\documentclass[12pt]{article}
\usepackage{latexsym}
\usepackage{graphicx}
\usepackage{enumitem}
\title{Bohmian mechanics for instrumentalists}
\author{Hrvoje Nikoli\'c \\
Theoretical Physics Division, Rudjer Bo\v{s}kovi\'{c} Institute, \\
P.O.B. 180, HR-10002 Zagreb, Croatia \\
{\normalsize e-mail: hnikolic@irb.hr} \\
\makebox[1in]{} \\
}
\date{\today}
\begin{document}
\maketitle
\begin{abstract}
We formulate Bohmian mechanics (BM) such that
the main objects of concern are macroscopic phenomena, 
while microscopic particle trajectories only play an auxiliary role. 
Such a formulation makes it easy to understand why BM always makes 
the same measurable predictions as standard quantum mechanics (QM), irrespectively of the details 
of microscopic trajectories. 
Relativistic quantum field theory (QFT) is interpreted as an effective long-distance theory that 
at smaller distances must be replaced by some more fundamental theory. 
Analogy with condensed-matter physics 
suggests that this more fundamental theory could have a form of non-relativistic QM, 
offering a simple generic resolution of an apparent conflict between BM and relativistic QFT.
\end{abstract}

\vspace{0.7cm}

\noindent
Keywords: Bohmian mechanics; relativistic QFT; effective theories; condensed matter physics 

\section{Introduction}

%
Quantum interpretations (see e.g. \cite{ghirardi,genovese_interp} for short reviews
and \cite{genovese_exp} for a review of their experimental status)
are frameworks for thinking about conceptual questions that cannot be
answered by the standard quantum formalism. One of the interpretations is
Bohmian mechanics (BM) \cite{bohm,book-bohm,book-hol,book-durr},
which
is usually (but not always \cite{oriols}) 
presented as a theory the main point of which is to offer 
a fundamental microscopic ontology - trajectories of elementary particles. 
For many physicists, especially instrumentalists, such a point of departure may not look intuitive 
and well motivated, the consequence of which is that BM is widely ignored or misunderstood 
in a wider physics community. Besides, BM is best understood as an interpretation of non-relativistic QM,
while its generalization to relativistic quantum field theory (QFT) is still an unsolved problem.
Various approaches to the Bohmian interpretation of relativistic QFT have been proposed 
(see e.g. \cite{book-hol,bell,durrcr2,struyve,nikQFTpilot,nikbook} and references therein),
but neither of those approaches seems completely satisfying.  

In this paper we offer a new instrumentalist-friendly formulation 
of BM in which the main objects of concern are macroscopic phenomena, especially readings 
of the measuring instruments, while microscopic ontology only plays a secondary auxiliary role. 
With such a reformulation it is much easier to understand intuitively why BM always makes 
the same measurable predictions as standard quantum mechanics (QM), irrespectively of the details 
of microscopic ontology. Furthermore, from the point of view that all currently 
well established physical theories are merely effective theories 
\cite{anderson,weinberg,effbook} 
describing only the phenomena 
at sufficiently large distances, it seems plausible to assume that relativistic quantum field theory (QFT) 
must be replaced by a completely different theory at smaller distances. 
By extending the ideas introduced in \cite{nikmin},
we use the analogy with condensed-matter physics 
to suggest that this more fundamental theory could have a form of non-relativistic QM, 
which offers a simple generic resolution of an apparent conflict between BM and relativistic QFT.

The paper is organized as follows. In Sec.~\ref{SECphilosophy} we present some philosophical 
preliminaries. In particular, we introduce and explain the concept of a perceptible -- 
a macroscopic phenomenon amenable to a direct human perception -- 
which is the central concept in our instrumental formulation of physics. 
In Sec.~\ref{SEC2} we formulate QM in terms of perceptibles, emphasizing the origin of 
an instrumental version of the Born rule. 
In Sec.~\ref{SECBohm} we introduce Bohmian mechanics as a natural completion of the 
formulation of QM in terms of perceptibles.
In Sec.~\ref{SEC5} we discuss how BM can naturally be made compatible with relativistic QFT
by arguing that relativistic QFT is not a fundamental theory. 
Finally, in Sec.~\ref{SEC6} we present a brief summary of our results.

\section{Philosophical preliminaries}
\label{SECphilosophy}

\subsection{Four basic notions in philosophy of physics}

Let us start with definitions of four basic concepts in philosophy of physics 
that we shall use in the rest of the paper.

{\it Ontology:} 
Things which are supposed to be there irrespective of (human) 
observations.

{\it Determinism:} 
The property that the future is completely determined by the past. 
Here ``determined'' can have two different meanings, namely (i) determined {\em in principle}
or (ii) determined {\em in practice}, e.g. by human-made computers with restricted computational power.
A prominent example is deterministic chaos, which is deterministic in principle but not in practice.
In the rest of the paper, unless explicitly stated otherwise, by ``determined'' we mean determined in principle. 
If determinism in that sense is right, the fundamental laws of physics are not probabilistic.
  
{\it Instrumentalism:} 
The stance that the main goal of physics is to predict and control 
the macroscopic phenomena, especially the outcomes 
of scientific instruments. 
In that sense, most physicists are (at least partly) instrumentalists.

{\it Instrumental interpretation of QM:} 
According to this interpretation, QM only prescribes probabilities of measurement outcomes.
Hence this interpretation is not deterministic in the practical sense.
However, it usually says nothing about
determinism in principle or about ontology, as long as it does not help 
to predict the macroscopic phenomena. 
For instance, it usually does not offer answers 
to questions such as ``Does particle have a position before one measures it?''. 

\subsection{The trouble with Bohmian mechanics}

Bohmian mechanics (BM) postulates that quantum particles are pointlike objects 
with deterministic trajectories. 
This postulate is usually motivated by the goal of prescribing a fundamental microscopic ontology. 
Namely, according to BM, a particle has a position even if one does not measure it.
Contrary to a widespread misconception, determinism is {\em not} the main goal of BM.
The determinism of BM is only a byproduct. 

Typical instrumentalists, however, do not care about ontology. 
Instrumentalists often say that ontology is not physics, but metaphysics. 
Hence instrumentalists usually do not find BM intuitive and well motivated. 
The consequence is that BM is widely ignored or misunderstood in a wider physics community.

The main goal of this paper is to {\it reformulate} BM  
such that it looks better motivated and more intuitive 
to a wider physics community, especially instrumentalists. 

\subsection{Three funny ``ble'' nouns in QM}

In the rest of the paper it will be crucial to understand the difference
between three words, the first of which is well known in QM,
the second of which is less known, and the third of which is completely new in QM.
Here we give brief explanations of their meaning.

{\it Observable:} 
In QM it is a noun (while in normal English it is an adjective). 
An observable is a hermitian operator in the Hilbert space.
It is related to a quantity that can be measured
in an actual experiment, but it is {\em not identical} 
to a quantity that can be measured in an actual experiment.
An operator in the Hilbert space is only a mathematical tool that theorists use
to {\em describe} the nature,
no experimentalist has ever seen an operator in the nature itself.  
As beautifully expressed by the instrumentalist Asher Peres \cite{peres}, 
``Quantum phenomena do not occur in a Hilbert space, they occur in a laboratory.''.

{\it Beable:} 
That word was coined by John Bell \cite{bell}. 
It means the same as ontology: stuff which is there irrespective of observation. 
The concept of a beable is central to Bohmians, but not to instrumentalists. 

With a goal to make BM more meaningful to instrumentalists, 
we find useful to introduce one additional concept, which we call perceptible.

{\it Perceptible:} 
In physics it is a noun (while in normal English it is an adjective). 
It is a thing or phenomenon amenable to {\it direct} human perception. 
Some examples of perceptibles are tables, chairs, the Moon, macroscopic instrument, 
click in the detector, the macroscopic picture of an atom produced by electron microscope, etc.  
Some examples of non-perceptibles are wave function, electron, photon, atom, etc. 

\subsection{More on perceptibles and beables}
\label{SEC2.4}

Since perceptible is a new concept that will be essential for understanding the rest of the 
paper, here we provide some additional explanations of that concept. 
In addition, we explain how the concept of beable can be interpreted from an instrumental
point of view.
 
The distinction between perceptibles and non-perceptibles is similar to 
the distinction between macroscopic and microscopic. 
In particular, all microscopic entities are non-perceptibles. 
However, a macroscopic entity does not necessarily need to be a perceptible.
For example, gravitational field and radio wave are macroscopic but not perceptibles. 

Non-perceptibles are {\it theoretical constructs} 
that help to explain and predict the properties of perceptibles. 
Some examples are given in Table 1.
In general, {\it to make a measurable prediction means to predict a property of a perceptible.}

\begin{table}[t]
\label{table1}
\begin{tabular}{|c|c|}
\hline
{\bf perceptible} & {\bf explained by non-perceptible} \\
\hline \hline
click in the photon detector & photon \\
picture of atom produced by electron microscope & atom, electrons \\
falling apple & gravitational field \\
music from the radio & radio wave \\
\hline
\end{tabular}
\caption{Examples of perceptibles and the corresponding non-perceptibles that explain the perceptibles.}
\end{table}

We said that a non-perceptible is a theoretical construct. However, when we say that
it is a theoretical construct, we do not say that it is not a beable. 
In fact, a beable is a theoretical construct itself; 
the claim that something is a beable really means that 
it is a beable {\em in a given theory}. 
For instance, point-like particles are beables in classical mechanics,
but not in classical field theory such as classical electrodynamics with continuous charge distributions. 
Hence, by saying that something is a beable, in this paper we do not necessarily claim that it 
exists in the reality.
A beable may exist according to a theory, but the theory does not need to be identical with the reality 
(the map is not the territory, as the proverb says).
Nevertheless, to intuitively understand a given theory, it may be useful to {\em imagine} that 
its beables exist in the reality, 
because that sometimes helps to {\em think} about the theory.
Hence we adopt an instrumental interpretation of a beable,
according to which it is just a useful thinking tool. 

We also emphasize that, like with macroscopic and microscopic, there is no strict border 
between perceptible and non-perceptible. 
For instance, is a one-cell microorganism macroscopic or microscopic? 
There is no unambiguous answer to that question.
Similarly, is perception of a one-cell microorganism by optical microscope direct or indirect? 
It must be direct to be called a perceptible, but there is no unambiguous border
between direct and indirect perception. A perception by a naked eye is clearly
direct, a perception by an electron microscope is clearly indirect,
but a perception by an optical microscope is somewhere in between.
Nevertheless, even though there is no strict border between micro and macro,
or between perceptibles and non-perceptibles, the concepts are useful.

By contrast, the property of being a beable is a binary property.
An object is either a beable or not, it 
cannot be somewhere in between.
One object can be more macroscopic than another, but one object cannot be more 
a beable than another. 

\section{Quantum theory of perceptibles}
\label{SEC2}

\subsection{All perceptibles can be reduced to macroscopic positions}
\label{SEC2.1}

As we said, all perceptibles are macroscopic. This means that they are large in the position space. 
Hence, when two perceptibles can be distinguished, it means that 
they can be distinguished by macroscopic positions of something. 

The simplest example is the measurement of spin. 
The spin is an observable, but it is not a perceptible. 
As discussed in many QM textbooks, spin can be measured by the Stern-Gerlach apparatus. 
In this case, the perceptible is a big dark spot on the screen of the apparatus.

Similar can be said about other, usually more sophisticated instruments,
which are either digital or analog.   
For analog instruments, the perceptible is a position of a macro pointer. 
For digital instruments, the perceptibles are positions of the lines that make a digit on the screen.

How about a click in the detector? 
The sound is determined by macroscopic oscillations of something, 
e.g. of a membrane of the speaker. 
This oscillation is nothing but a macroscopic position as a function of time. 
Hence a click in the detector can also be reduced to macroscopic positions.

What about senses such as color, taste or smell? 
They are created in the eye, tongue or nose respectively (and interpreted by brain),
by stimulation of a nerve. For the creation of a sense it is not so much important 
{\em what} stimulates the nerve (what kind of light or chemical).
It is much more important {\em which} nerve is stimulated.
For instance, the electro-stimulation 
of the sweet nerve creates the illusion of sweetness, without having any 
sweet chemical. Since different nerves have different macroscopic positions,
it means that even senses such as color, taste and smell
can be reduced to macroscopic positions.

\subsection{The origin of Born rule in QM}
\label{SECbornrule}

The Born rule is usually presented as a {\em postulate} that,
in any Hilbert-state basis $\{|k\rangle\}$, the probability is given by 
\begin{equation}\label{born1}
 p_k=|\langle k|\psi\rangle|^2 .
\end{equation}
However, it is not really necessary to postulate it in all possible bases. 
It is sufficient to postulate it only in the position basis,
because the instrumental form of the general Born rule (\ref{born1}) can be {\em derived}  
from the Born rule in the position space.

To derive the instrumental form of the general Born rule (\ref{born1}), we start from the crucial insight that 
what we need are probabilities of perceptibles, such as the probability that a detector will click. 
The results of Sec.~\ref{SEC2.1} then imply that probabilities of perceptibles must be 
computed in the position space. 
However, since there is no strict border between 
perceptible and non-perceptible,  
it seems natural to compute {\em all} probabilities in the position space.
In this way, the position basis becomes a preferred basis not only for macroscopic degrees of freedom,
but even for the microscopic ones.

To understand it in more detail, suppose that we measure a microscopic observable $\hat{K}$ with eigenstates $|k\rangle$. 
For simplicity, we assume that $\hat{K}$ has a non-degenerate spectrum (in Sec.~\ref{SECgenmeas}
we shall discuss more general measurements). The microscopic observable can be written in the spectral form 
\begin{equation}
\label{K}
 \hat{K} = \sum_k k |k\rangle\langle k| .
\end{equation}
To describe the measurement, we also need to describe 
the macroscopic apparatus, that is the perceptible. The macroscopic apparatus can be described 
by its microscopic quantum state $|A\rangle$.
If the initial state of the measured system is $|k\rangle$ and 
if the initial microscopic state of the apparatus is $|A_0\rangle$,
then the interaction between the measured system and the apparatus induces a unitary transition
of the form
\begin{equation}\label{meas1}
 |k\rangle |A_0\rangle \rightarrow |k'\rangle |A_k\rangle .
\end{equation}
Here we allow that $|k'\rangle$ may be different from $|k\rangle$, which means that we consider
measurements slightly more general than projective measurements. (In Sec.~\ref{SECgenmeas}
we shall discuss even more general measurements.)
We require that different measurement outcomes are {\em macroscopically distinguishable},
for otherwise the interaction cannot be interpreted as a measurement.
Hence the apparatus wave functions have a negligible overlap in the multi-position space, i.e.
\begin{equation}\label{nooverlap}
 A_{k_1}(\vec{x}) A_{k_2}(\vec{x}) \simeq 0 \;\;\;\; {\rm for} \;\;\;\;  k_1\neq k_2 .
\end{equation}
Here $A_k(\vec{x}) \equiv \langle \vec{x}|A_k \rangle$, where
\begin{equation}
\vec{x} \equiv  ({\bf x}_1,\ldots,{\bf x}_n) , 
\end{equation}
$n$ is the number of particles constituting the apparatus, and the wave functions are normalized as 
\begin{equation}
\int d\vec{x} \, |A_{k}(\vec{x})|^2 =1 , 
\end{equation}
with $d\vec{x}\equiv  d^{3n}x$.

Similarly, if the initial state of the measured system is in a superposition 
\begin{equation}\label{psi}
|\psi\rangle = \sum_k c_k |k\rangle ,
\end{equation}
where 
\begin{equation}
\label{ck}
 c_k=\langle k|\psi\rangle ,
\end{equation}
then (\ref{meas1}) generalizes to
\begin{equation}\label{meas2}
 |\psi\rangle |A_0\rangle \rightarrow \sum_k c_k |k'\rangle |A_k\rangle .
\end{equation}
A more realistic analysis includes also the environment, so (\ref{meas2}) further 
generalizes to
\begin{equation}\label{meas3}
 |\psi\rangle |A_0\rangle |E_0\rangle \rightarrow \sum_k c_k |k'\rangle |A_k\rangle |E_k\rangle \equiv |\Psi\rangle .
\end{equation}
Hence the state after the measurement can be written in the final form 
\begin{equation}\label{meas4}
|\Psi\rangle = \sum_k c_k |A_k\rangle |R_k\rangle ,
\end{equation}
where the apparatus state $|A_k\rangle$ describes the perceptible and $|R_k\rangle = |k'\rangle|E_k\rangle$
describes all the rest. This state can be written in the multi-position representation as
\begin{equation}\label{meas4.1}
 \Psi(\vec{x},\vec{y})=\sum_k c_k A_k(\vec{x}) R_k(\vec{y}) ,
\end{equation}
where $\vec{y}$ are positions of all particles except those of the apparatus.
Hence the Born rule in the multi-position space 
\begin{equation}\label{born2}
\rho(\vec{x},\vec{y}) =|\Psi(\vec{x},\vec{y})|^2 
\end{equation}
gives the probability density
\begin{equation}\label{born2.1}
\rho(\vec{x},\vec{y}) \simeq \sum_k |c_k|^2 |A_k(\vec{x})|^2 |R_k(\vec{y})|^2 ,
\end{equation}
where (\ref{nooverlap}) has been used. Therefore the probability density for positions of
the apparatus particles is given by the marginal probability density
\begin{equation}
\label{rhox}
\rho^{\rm (appar)}(\vec{x})=\int d\vec{y} \, \rho(\vec{x},\vec{y})
\simeq \sum_k |c_k|^2 |A_k(\vec{x})|^2 .
\end{equation}
Now let the support of $A_{k}(\vec{x})$ be the region of $\vec{x}$-space in which $A_{k}(\vec{x})$ 
is not negligible. Then the probability to find the apparatus particles in the support of $A_{k}(\vec{x})$ is
\begin{equation}
\label{pk}
 p_{k}^{\rm (appar)}=\int_{\rm supp \;\; A_{k}} d\vec{x} \,  \rho^{\rm (appar)}(\vec{x}) \simeq |c_k|^2 .
\end{equation}
Recalling (\ref{ck}), we see that (\ref{pk})
coincides with the Born rule (\ref{born1}) in arbitrary $k$-space, which finishes the derivation of 
the general Born rule (\ref{born1}) from the Born rule (\ref{born2}) in the position space. 

\subsection{Generalized measurements}
\label{SECgenmeas}

To emphasize the importance of Eq.~(\ref{meas4}), we refer to it as 
the {\em master formula} of quantum measurement
and write it in a slightly more general form
\begin{equation}\label{master}
|\Psi\rangle = \sum_l \tilde{c}_l |A_l\rangle |R_l\rangle .
\end{equation}
In the master formula (\ref{master})
(not to be confused with the master {\em equation} in the theory of quantum decoherence \cite{schloss}),
$|A_l\rangle$ denotes the microscopic state of the perceptible, $|R_l\rangle$ represents 
the microscopic state of all the rest, and $\tilde{c}_l$ are some coefficients obeying $\sum_l|\tilde{c}_l|^2=1$.
Eq.~(\ref{master}) is a generalization of (\ref{meas4}) in the sense that 
the coefficients $\tilde{c}_l$ do not need to be equal to $c_k$ given by (\ref{ck}). 
In general, the coefficients $\tilde{c}_l$ depend not only on the initial state (\ref{psi})
of the measured system, but also on details of its interaction   
with the apparatus and the environment.

In the derivation of the master formula (\ref{meas4})
in Sec.~\ref{SECbornrule}, the label $k$ had a double meaning: 
\begin{enumerate}[label=(\roman*)]
\item
Eigenstates $|k\rangle$ of an observable $\hat{K}$ with a non-degenerate spectrum. 
\item
Label of distinct perceptibles. 
\end{enumerate}

For more general measurements, however, the label $l$ in (\ref{master})
does not need to have the meaning (i).  
Examples where the meaning (i) cannot be applied include measurement of an observable 
with a degenerate spectrum, measurement of the photon position, measurement of time, etc. 
Nevertheless, all such generalized (not necessarily projective) measurements can be described by the POVM formalism 
\cite{peres,nielsen-chuang,muynck,audretsch,schumacher,laloe,witten} 
and the Neumark's theorem \cite{peres} provides that any POVM measurement can be reduced 
to a projective measurement in some larger Hilbert space. 
Physically, this means that the master formula (\ref{master}) with the meaning of $l$ given by (ii) 
is valid for {\em any} measurement with clearly distinguishable outcomes. 
Identifying the larger Hilbert space with the product ${\cal H}_A\otimes{\cal H}_R$
of the apparatus Hilbert space ${\cal H}_A$ and the Hilbert space ${\cal H}_R$ of the rest,
the relevant projectors in the larger Hilbert space are $|A_l\rangle\langle A_l| \otimes 1$.
So even if there is no microscopic observable (\ref{K}), one can always interpret the measurement 
as a projective measurement of the apparatus observable
\begin{equation}
\label{K2}
 \hat{L}^{\rm (appar)} = \sum_l l  |A_l\rangle\langle A_l|.
\end{equation}
Hence, by repeating the analysis described by Eqs.~(\ref{meas4.1})-(\ref{pk}), we conclude that
the instrumental form of the Born rule 
\begin{equation}
 p_{l}^{\rm (appar)}\simeq |\tilde{c}_l|^2
\end{equation}
is always true, whenever the probability density in the multi-position space
is given by $\rho(\vec{x},\vec{y};t) =|\Psi(\vec{x},\vec{y},t)|^2$.

\section{Bohmian mechanics}
\label{SECBohm}

\subsection{Motivation for BM}

In contrast with more traditional approaches to BM,
in our instrumental approach to BM we do {\em not} start from postulating particle trajectories. 
Instead, our main postulate is this: 
\newtheorem{coms}{Axiom}
\begin{coms}
\label{a1}
All perceptibles are beables.
\end{coms}
Loosely speaking, it says that the Moon is there even if nobody observes it.
The opposite would be that the Moon is only in our mind,
which few physicists would find appealing.
Hence the Axiom \ref{a1} is motivated by common sense, 
but we stress that it is impossible to prove or disprove this axiom by the scientific method. 
This axiom is nothing but a useful thinking tool. It is useful because it is mentally 
hard (though not impossible \cite{niksolip}) to think the opposite in a logically consistent manner.  
As we shall see, most of the motivation for BM arises from this common sense axiom.

Before discussing how that leads to BM, let us note that the famous Bell theorem \cite{bell} 
can be expressed in the language of perceptibles as follows: 
{\it If perceptibles described by QM are beables, then those perceptibles obey non-local laws.}  
That is, if the correlated, yet spatially separated, quantum measurement outcomes 
are there even before a single local observer detects the correlation, 
then the measurement outcomes are governed by non-local laws \cite{mermin,tumulka}. 
Note that the Bell theorem expressed in that form avoids the notion of ``hidden variables'' 
and does not depend on the assumption of determinism. 

As we explained, a perceptible is determined by the microscopic positions $\vec{x}=({\bf x}_1,\ldots,{\bf x}_n)$ 
of the apparatus particles. In principle one might conceive that the microscopic positions are not beables,
even though the macroscopic position of the perceptible is a beable. 
However, as we explained in Sec.~\ref{SEC2.4}, something can be less macroscopic than something else, 
but it cannot be less a beable than something else. Hence, to conceive that only macroscopic positions are beables,
one would need to introduce a strict border between micro and macro positions, 
which would look {\em ad hoc} and conceptually complicated.
Therefore, to avoid such complications, we propose that {\em all} $\vec{x}$ are beables, which is
the simplest possibility compatible with the Axiom \ref{a1}.
Similarly, as there is no strict border between perceptibles and non-perceptibles, 
we propose that positions $\vec{y}$ of all the rest are beables too. 

In Sec.~\ref{SEC2} we have derived the quantum mechanical Born rule in arbitrary $k$-space 
from the Born rule in the position space. 
This means that {\em any} theory, for which the probability density in position space is
\begin{equation}
\rho(\vec{x},\vec{y};t) =|\Psi(\vec{x},\vec{y},t)|^2 ,
\end{equation}
has the same measurable predictions as QM. 

This is valid even for generalized measurements discussed in Sec.~\ref{SECgenmeas}.
As an example, consider measurement of time. 
It is well known that there is no time operator $\hat{K}=\hat{T}$ with eigenstates $|k\rangle=|t\rangle$ 
\cite{timeqm,nikmyths}. 
Nevertheless it is not a problem for measurement of time, because in the master formula (\ref{master})
written as
\begin{equation} 
\label{master2} 
|\Psi\rangle = \sum_l \tilde{c}_l |C_l\rangle |R_l\rangle,
\end{equation}
the label $l$ just labels the distinct macroscopic positions of the macroscopic clock pointer 
with microscopic clock states $|C_l\rangle$.  

So far, in this section we have found motivation for two requirements. The first requirement is that
perceptibles are beables, which was motivated by common sense.
The second requirement is that $\rho(\vec{x},\vec{y};t) =|\Psi(\vec{x},\vec{y},t)|^2$, which was motivated by QM.
What kind of theory can satisfy both requirements?

A simple theory that satisfies both requirements is a theory in which 
all positions $\vec{q}=(\vec{x},\vec{y})$ are {\em beables} and {\em random}. 
This looks almost like standard QM, except that standard QM does not insist that $\vec{q}$ are beables. 
However, such a theory in which the positions $\vec{q}$ are random  
does {\em not explain} the Born rule for $\vec{q}$. 
In this theory, the Born rule for $\vec{q}$ is still {\em postulated}.

Can we do better? Can we {\em explain} the Born rule for $\vec{q}$? 
The requirement that $\vec{q}$ is a beable implies that it has a value $\vec{Q}(t)$ at each time $t$. 
In principle $\vec{Q}(t)$ could be a stochastic (that is, non-deterministic) function of time $t$. 
However, $\vec{Q}(t)$ must be compatible with $\rho(\vec{q};t) =|\Psi(\vec{q},t)|^2$, 
and the Schr\"odinger equation implies that $|\Psi(\vec{q},t)|^2$ is a deterministic function of $t$. 
This suggests (but not proves) that $\vec{Q}(t)$ could be deterministic too.

To summarize, the requirement that perceptibles are beables suggests that $\vec{Q}(t)$ exists,
while the requirement of compatibility with the Schr\"odinger equation suggests that $\vec{Q}(t)$ is deterministic. 
This is the motivation for introducing BM.

\subsection{Construction of BM}
\label{SEC4.2}

How can a deterministic law for $\vec{Q}(t)$ be compatible
with the probability density $\rho(\vec{q};t) =|\Psi(\vec{q},t)|^2$? 
The compatibility condition is that $\vec{Q}(t)$ is determined by a law of the form
\begin{equation}\label{BohmQ}
 \frac{d\vec{Q}(t)}{dt}=\vec{v}(\vec{Q}(t),t) ,
\end{equation}
where $\vec{v}(\vec{q},t)$ is a function that satisfies the continuity equation 
\begin{equation}\label{cont}
 \frac{\partial|\Psi|^2}{\partial t}+\vec{\nabla}(|\Psi|^2\vec{v})=0 .
\end{equation}
Namely, if $\rho(\vec{q};t_0) =|\Psi(\vec{q},t_0)|^2$ for initial $t_0$, 
then the continuity equation implies that $\rho(\vec{q};t) =|\Psi(\vec{q},t)|^2$ for all $t$. 

In fact, the continuity equation is analogous \cite{book-hol,book-durr} to the Liouville equation 
in classical statistical mechanics. 
This means that the probability density $\rho(\vec{q};t) =|\Psi(\vec{q},t)|^2$ 
corresponds to a {\em quantum equilibrium}, which can be explained 
even without assuming the initial condition $\rho(\vec{q};t_0) =|\Psi(\vec{q},t_0)|^2$. 
Similarly to classical statistical mechanics \cite{tolman,ehrenfest},
there are two different approaches to the explanation of quantum equilibrium 
from the continuity equation. One approach is based on a $H$-theorem \cite{valentini}
and the other approach is based on typicality \cite{durr_absunc}.
For a review of both approaches see \cite{norsen}.

So is there such $\vec{v}=({\bf v}_1,\ldots, {\bf v}_N)$,
with $N$ being the number of particles,
that satisfies the continuity equation (\ref{cont})?
In non-relativistic QM the answer is yes, because
it is well known that the Schr\"odinger equation itself 
\begin{equation}\label{sch}
 \hat{H}\Psi=i\hbar\partial_t\Psi
\end{equation}
with
\begin{equation}\label{sch2}
 \hat{H}=\sum_{a=1}^N \frac{\hat{\bf p}_a^2}{2m_a} +V(\vec{q})
\end{equation}
implies a continuity equation of the form (\ref{cont}) with
\begin{equation}\label{Bohmv}
 {\bf v}_a =  \frac{-i\hbar}{2m_a} \,
\frac{\Psi^* \!\stackrel{\leftrightarrow\;}{ \mbox{\boldmath $\nabla$}_a }\!  \Psi}
       {\Psi^*\Psi} =
\frac{ {\rm Re} (\Psi^*\hat{\bf v}_a \Psi)}{\Psi^*\Psi} ,
\end{equation}
where $\hat{\bf v}_a=\hat{\bf p}_a/m_a$ is the velocity operator and
$\hat{\bf p}_a=-i\hbar \mbox{\boldmath $\nabla$}_a$ is the momentum operator. 

The quantity $|\Psi|^2\equiv \Psi^*\Psi$ is the probability density for particles without spin.
Similarly, (\ref{Bohmv}) also describes particles without spin.
To generalize those expressions to particles with spin, all one has to do is to  
make the replacement
\begin{equation}
\Psi^*\cdots\Psi\rightarrow \Psi^{\dagger}\cdots\Psi=\sum_{\alpha}\Psi_{\alpha}^*\cdots\Psi_{\alpha} ,
\end{equation}
where $\sum_{\alpha}$ represents the sum over all spin indices.

The results above show that BM works for non-relativistic QM. 
But we stress that the law (\ref{BohmQ}) with (\ref{Bohmv}) is non-local when 
the wave function is in the entangled state.
The velocity of any particle at time $t$ depends on the positions of all particles
at the same time $t$, no matter how far the other particles are. 
We also stress that this non-local interaction is the only interaction between Bohmian point-like particles.
When the wave function is not entangled, then Bohmian point-like particles do not interact with each other at all.
In this sense, Bohmian particles have {\em only} non-local interactions.
However, one should distinguish this interaction between Bohmian point-like particles from the interaction 
described by the Hamiltonian in the Schr\"odinger equation. The latter describes the evolution of the 
wave function, which should be distinguished from the evolution of Bohmian particle positions.
Hence the fact that Bohmian particles have only non-local interactions is not in contradiction 
with the possibility of having a Hamiltonian with local interactions.

\subsection{Robustness of long distance physics}
\label{SECrobust1}

In physics, there is a general rule of thumb that says that 
{\em the laws of long distance physics 
do not depend on details of small distance physics.}  
There are many examples of that rule. 
For instance, fluid mechanics and thermodynamics do not depend 
on details of atomic physics. (One does not need to know much about
atomic physics to be an expert in fluid mechanics or thermodynamics.) 
Similarly, atomic physics does not depend on details of nuclear physics. 
Likewise, nuclear physics does not depend on details of quarks. 
And finally quantum chromodynamics, the standard theory of quarks and gluons,
does not depend on details (or even validity) of string theory. 

This intuitive rule can be formalized more generally by the Wilson renormalization theory
(see e.g. \cite{peskin}). According to the Wilson renormalization theory,
the long distance physics is obtained from more fundamental microscopic theory 
by {\em integrating out} the small distance degrees of freedom. 

This general rule helps to understand why BM has the same measurable 
predictions as standard QM, which we discuss in more detail in the next
subsection.

\subsection{Robustness of measurable predictions by BM}

Similarly to the general rule in physics discussed in Sec.~\ref{SECrobust1}, 
the perceptibles in BM do not depend on details of particle trajectories. 
This is seen from Eqs.~(\ref{rhox}) and (\ref{pk}), which say that probability  
of a perceptible is obtained by {\em integrating out} over all microscopic positions
\begin{equation}
 p_{l}^{\rm (appar)}=\int_{\rm supp \;\; A_{l}} d\vec{x} \int d\vec{y} \, |\Psi(\vec{x},\vec{y})|^2 .
\end{equation}
Intuitively, it says that the precise particle positions are not very much important 
to make measurable predictions. It is important that particles have {\em some} positions 
(for otherwise it is not clear how can a perceptible exist), but it is much less important
what exactly those positions are.
That is why BM (with trajectories) makes the same measurable 
predictions as standard QM (without trajectories). 

It is extremely important not to overlook the general idea above that the precise particle positions
are not essential
%
(unless some special conditions are met, which we discuss at the end of this section). 
For otherwise, one can easily make a {\em false} ``measurable prediction'' out of BM that seems to differ from 
standard QM, when in reality there is no such measurable prediction.
The general recipe for making such a false ``measurable prediction'' out of BM is  
to put too much emphasis on trajectories and ignore the perceptibles. 
A lot of wrong ``disproofs of BM'' of that kind are published in the literature.

By a peer pressure of making new measurable predictions out of BM,
even distinguished Bohmians sometimes fall into this trap.
For instance, some try to make new measurable predictions of arrival times
by computing the arrival times of microscopic BM trajectories 
(see e.g. \cite{leavens,durrtime}). 
However, the microscopic trajectories are not perceptibles,
so the arrival times obtained from microscopic BM trajectories may be rather deceptive 
from a measurable point of view. 
To make a measurable prediction, one must first specify how exactly 
the arrival time is {\em measured} \cite{steinberg}, which requires a formulation of the problem
in terms of a perceptible. When the problem is formulated in that way,
BM makes the same measurable predictions as standard QM, despite the fact
that there is no time operator in standard QM (recall also the discussion around Eq.~(\ref{master2})).

Another example of Bohmian computations that may look deceptive from a measurable
point of view are computations of the gravitational field $g_{\mu\nu}^{\rm (Bohm)}({\bf x},t)$ in Bohmian quantum 
cosmology (see e.g. \cite{p-n-struyve} and references therein).
Even though the gravitational field is macroscopic, we recall from Sec.~\ref{SEC2.4} 
that it is not a perceptible. Moreover, $g_{\mu\nu}^{\rm (Bohm)}({\bf x},t)$
is not a classical gravitational field, so analysis of a measurable effect
should involve some version of the quantum master formula (\ref{master}). 
As long as computations in Bohmian quantum cosmology do not discuss consequences of (\ref{master}), 
it is not clear what, if anything, 
is the meaning of $g_{\mu\nu}^{\rm (Bohm)}({\bf x},t)$ from a phenomenological point of view.

BM is deterministic, so why cannot it make deterministic predictions  
of measurement outcomes? 
It is because of the quantum equilibrium \cite{valentini,durr_absunc}, 
which is most easily understood through the analogy with thermal equilibrium. 
In the full thermal equilibrium, macroscopic changes can only happen 
due to rare statistical fluctuations. In principle, 
those fluctuations are in classical statistical physics governed by deterministic laws. 
Nevertheless, owing to a large number of particles,
those deterministic laws cannot be used to make deterministic predictions 
{\em in practice}.
In practice, thermodynamics makes deterministic predictions of macroscopic changes 
only when the full system is {\em not} in a thermal equilibrium. 

In addition, we stress that one does not need to explain why a system is in the equilibrium,
for equilibrium is the most probable state. 
It is rather the {\em absence} of equilibrium that needs an explanation. 
For instance, it is still not clear why the Universe is not in a thermal equilibrium.  
On the other hand, the fact that the Born rule is consistent with all existing experiments
is evidence that the Universe is in the quantum equilibrium.

Why cannot BM trajectories be observed? Or more precisely, why cannot
a single measurement reveal a Bohmian particle position  
with a precision better than the spatial width of the wave function? 
This is not only because Bohmian positions are not perceptibles;
after all atom positions are also not percetibles, yet electron microscope can be used 
to observe atom positions.  
The true reason why Bohmian positions cannot be observed 
with a precision better than the spatial width of the wave function
is because there are no local interactions (in the sense explained in Sec.~\ref{SEC4.2}) 
between BM particles.
To make an analogy, trying to observe a Bohmian trajectory 
is like trying to observe the Moon's trajectory by watching tides. 
Classical gravity is a long range force, so the observation of effect on B 
caused by A does not directly reveal the position of A. 
That is why we cannot observe the Moon's trajectory by watching tides.
%

%
Finally let us note that, even though the quantum equilibrium is a natural hypothesis
consistent with observations, it is not a logical necessity. Deviations 
from quantum equilibrium are in principle possible, in which case the precise Bohmian particle positions
may be important and Bohmian mechanics may lead to measurable predictions
that differ from standard QM, as proposed e.g. in 
\cite{valentini-noeq,golshani}. The proposal in \cite{golshani} has been ruled out experimentally in \cite{brida}.

\section{Beyond relativistic QFT}
\label{SEC5}

\subsection{What particles is BM about?}
\label{SEC5.1}

So far we were talking about particle trajectories, but
we did not specify what kind of particles we are talking about. 
Is it atoms? Or is it smaller particles such as protons and neutrons? 
Or is it more fundamental particles like electrons, quarks, photons and Higgs? 
Or perhaps quasiparticles (collective excitations), like phonons? 
As we stressed several times, the predictions on perceptibles do not depend much on those details. 
Yet the details are important for their own sake.

Take for example a phonon.
The phonon trajectory is certainly not a beable because we {\em know} that 
one phonon is a collective motion of many atoms. 
Photons and electrons, on the other hand, are usually viewed as 
fundamental particles, rather than quasiparticles.
But do we actually {\em know} that photon or electron is not a collective excitation of 
some more fundamental degrees of freedom? 
We stress that we do {\em not know} that. 

Let us explain.
Theories which serve as good approximations at longer distances, 
but not at smaller distances, are called {\em effective theories} \cite{effbook}. 
The theory of phonons is certainly an effective theory. 
There is some theoretical evidence that the Standard Model of ``elementary particles'' 
is an effective theory too \cite{weinberg,effbook}. 
%
One evidence is the fact that the Standard Model has a lot of parameters (masses, coupling constants and mixing angles)
which are extracted from experiments but cannot be explained by the Standard Model itself,
suggesting that there is an unknown more fundamental theory that should explain them.
Another evidence is the fact that the Standard Model is a continuous quantum field theory (QFT) 
with ``ultraviolet'' short-distance divergences, the consequence of which is that it cannot be 
formulated in a mathematically rigorous way, which suggests that there should be a more fundamental 
theory that does not suffer from such mathematical problems.
Yet another evidence is the fact that the Standard Model does not contain quantum gravity,
which is a strong indication that the Standard Model is incomplete.    
It is not known what the fundamental theory behind the Standard Model is,
but one viable possibility is that 
the ``elementary particles'' like electrons, quarks, photons, etc.
are in fact collective excitations. 
Collective excitations of what? Of some truly elementary particles. 
What those truly elementary particles are? 
We do not know, because we still do not have the theory of everything. 
We can only say that this hypothetic truly elementary particles are for 
photons (and other Standard Model ``elementary particles'') what the atoms are for phonons.

How is that relevant to BM? The point is that Bohmian particle trajectories can only be beables 
for the truly elementary particles, whatever those truly elementary particles are.
(For an illustration of that idea see also Sec.~\ref{SECexample}.)
Hence it seems very likely that
{\em BM trajectories are {\bf not} beables for the Standard Model 
``elementary particles'' like electrons, quarks, photons and Higgs.}

\subsection{Bypassing relativistic QFT}

In Sec.~\ref{SEC4.2}
we found an explicit construction of BM for non-relativistic QM. 
But what about relativistic QFT?  

The ``elementary particles'' such as electrons, photons etc. 
are described by relativistic QFT. 
But in Sec.~\ref{SEC5.1} we argued that we do not need BM trajectories for them. 
The BM trajectories are only needed for truly elementary particles. 
We do not know yet what the theory of those truly elementary particles is,
but in principle it is possible that the truly elementary particles
are not described by relativistic QFT. 
In fact, there is a good heuristic argument for that. 
For truly fundamental particles it makes sense to assume that they cannot be created and destroyed.
On the other hand, it is known that relativistic QM and QFT naturally lead to particle creation and destruction
(see e.g. \cite{BD1,BD2}).  
Hence it seems reasonable to assume that the truly fundamental particles, if they exist, 
are described by non-relativistic QM. 
If so, then BM bypasses relativistic QFT.
 
This can also be interpreted as a generic {\em measurable prediction} of BM. 
The simplest formulation of BM, described in Sec.~\ref{SEC4.2},
requires that the most fundamental 
degrees are described by non-relativistic QM. 
This means that the simplest version of BM predicts that at some very small distances 
(not yet amenable to our current experimental technologies such as those in Large Hadron Collider at CERN) 
we should find new particles the observable properties of which do not obey Lorentz invariance. 
This generic prediction differs e.g. from generic predictions of string theory \cite{GSW}.

\subsection{How could it be that non-relativistic QM is fundamental 
and that relativistic QFT is only an approximation?}

It is usually considered that relativistic QFT is fundamental, 
while non-relativistic QM is only an approximation. 
By contrast, we propose that the opposite is the case. How could that be? 
The basic idea is presented in most textbooks on condensed-matter physics. 

First, to avoid terminological confusion, let us say that by ``relativistic'' we mean 
a property of any theory (fundamental or not) that it is invariant under Lorentz transformations with some
constant speed parameter $\tilde{c}$. Explicitly, for a boost in the $x$-direction, 
those Lorentz transformations are
\begin{equation}\label{lorentz}
x'=\frac{x-vt}{\sqrt{1-v^2/\tilde{c}^2}}, \;\;\; y'=y, \;\;\; z'=z, \;\;\; 
t'=\frac{t-vx/\tilde{c}^2}{\sqrt{1-v^2/\tilde{c}^2}} .
\end{equation} 
The usual notion of relativistic theory is a special case, in which the speed parameter
is fixed to be the speed of light $\tilde{c}=c$.

Now consider sound. Sound waves satisfy the wave equation
\begin{equation}
\label{sound}
 \frac{1}{c_s^2} \frac{\partial^2\psi}{\partial t^2} -\nabla^2 \psi =0 .
\end{equation}
This equation is Lorentz invariant under Lorentz transformations 
in which the speed parameter is the speed of sound $\tilde{c}=c_s$. 
Eq.~(\ref{sound}) is a valid approximation only at distances much larger than interatomic distances,
so the Lorentz invariance is an approximation valid only at those large distances. 
The sound wave is propagated through the medium made of atoms,
which means that the Lorentz invariant wave equation is derived from non-relativistic motion of atoms.
The medium can be thought of as the ``ether'' for sound waves. 
If one observed {\em only} the sound and nothing else, 
it would look as if there was no ``ether'' for sound and
the Lorentz invariance of (\ref{sound}) would look like a fundamental law.
In reality, we also observe phenomena other than sound, 
which is how we know that the Lorentz invariance of (\ref{sound}) is not fundamental.
 
The classical wave equation (\ref{sound}) can be ``quantized'', which leads to a
quantum theory of sound (see e.g. \cite{simons}).
First quantization of $\psi$ leads to QM of a single phonon.  
Second quantization of $\psi$ leads to QFT of phonons. 
Those are standard tools in condensed-matter physics. 
They are derived from non-relativistic QM of atoms, where each atom is thought of as an 
object made of a pointlike nucleus surrounded by electrons. 
In this way, the creation and destruction of relativistic phonons is derived 
from a non-relativistic theory of a fixed number of atoms. 

By analogy \cite{nikmin}, it is conceivable that all relativistic ``elementary particles'' of the Standard Model 
(photons, electrons, etc, which are relativistic with $\tilde{c}=c$) can be derived, in a similar way, from some hypothetic 
more fundamental non-relativistic particles. 
If so, then the world looks ``fundamentally'' relativistic only because 
we do not yet see those more fundamental degrees. 

Such a theory can be thought of as a neo-Lorentzian ether theory.
(For other theories of that kind in modern literature see e.g. \cite{nielsen,jacobson,horava}.)
The 19th century Lorentzian ether theory \cite{ether}
was ruled out by the famous Michelson-Morley experiment,
which ruled out the possibility that the Earth 
moves {\em through} the ether.
What we propose here is that the Earth (and everything else) 
is {\em made of} ether.
No experiment so far ruled out that possibility, so such a neo-Lorentzian ether theory
is a viable possibility.

In the literature there are many explicit models in which various qualitative properties of 
the Standard Model of ``elementary particles'' are derived from 
condensed-matter systems. For explicit models of that kind we refer readers to the books
\cite{volovik,wen} and references therein.
A model that could reproduce {\em all} quantitative details of the Standard Model is still 
not known, but the broadness and flexibility of the condensed-matter framework 
is a reason for optimism. In that sense, the condensed-matter framework in a search for a
more fundamental theory is similar to string theory, as string theory is also a broad and flexible framework
within which a model that could reproduce all quantitative details of the Standard Model
has not yet been found \cite{string_phen}.

\subsection{Example: A Phonon and its Bohmian interpretation}
\label{SECexample}

Let us illustrate the general qualitative ideas above by an example. 
Consider a crystal lattice made of $N$ atoms with positions
\begin{equation}
 \vec{q}=({\bf q}_1,\ldots, {\bf q}_N) .
\end{equation}
The wave function $\Psi(\vec{q},t)$ satisfies the non-relativistic Schr\"odinger equation
(\ref{sch})-(\ref{sch2}),
where $V(\vec{q})$ describes the inter-atomic interactions.
Let $\Psi_{\bf p}(\vec{q},t)$ be the solution corresponding to one acoustic phonon 
with momentum ${\bf p}$. 
Then the most general 1-phonon solution is
\begin{equation}
 \Psi(\vec{q},t)=\sum_{\bf p} c_{\bf p} \Psi_{\bf p}(\vec{q},t) .
\end{equation}
In the abstract Hilbert space this state can be written as 
\begin{equation}
|\Psi(t)\rangle = \sum_{\bf p} c_{\bf p} |\Psi_{\bf p}(t)\rangle ,
\end{equation}
which can also be represented by a 1-quasiparticle wave function 
\begin{equation}
 \psi({\bf x},t)=\sum_{\bf p} c_{\bf p} e^{-i[\omega({\bf p})t-{\bf p}\cdot{\bf x}] } .
\end{equation}
Here $\omega({\bf p})= c_s |{\bf p}|$ is the Lorentz invariant 
dispersion relation for acoustic phonons and we use units $\hbar=1$.
The 1-quasiparticle wave function $\psi({\bf x},t)$ satisfies wave equation
\begin{equation}
 \frac{1}{c_s^2} \frac{\partial^2\psi}{\partial t^2} -\nabla^2 \psi =0 .
\end{equation}

Now we want to understand how this system can be understood from a Bohmian point of view.
To illustrate the general idea that microscopic details of BM depend on what is
considered fundamental, we consider four different versions of Bohmian interpretation.

{\em Bohmian interpretation 1}: 
The existence of the phonon wave function $\psi({\bf x},t)$ suggests 
that one can introduce a phonon Bohmian trajectory ${\bf X}(t)$.  
It makes sense if one imagines that phonon is a fundamental particle.
Few physicists would take such a picture seriously.

{\em Bohmian interpretation 2}: 
Now consider a somewhat more serious version of BM. 
This version denies the existence of the phonon trajectory ${\bf X}(t)$, 
but the wave function $\Psi(\vec{q},t)$ of $N$ atoms suggests that atoms 
have Bohmian trajectories ${\vec Q}(t)$. 
It makes sense if one imagines that atoms are fundamental.
Within this interpretation one can identify three different wave-like objects:
\begin{itemize} 
\item The 1-phonon wave function $\psi({\bf x},t)$, which is relativistic and not fundamental.  
\item The multi-atom wave function $\Psi(\vec{q},t)$, which is non-relativistic and fundamental. 
\item The collective motion of atoms ${\vec Q}(t)$, which is non-relativistic and fundamental.
\end{itemize} 

{\em Bohmian interpretation 3}:
An even more serious version of BM denies the existence of atom trajectories ${\vec Q}(t)$, 
because the atoms are not fundamental, but made of ``elementary particles'' such as quarks and electrons.
Hence this version of BM accepts the existence of quark and electron Bohmian trajectories
${\vec Q}_{\rm quarks\; \& \;electrons}(t)$. 
It makes sense if one imagines that quarks and electrons are fundamental.
However, it requires a relativistic version of BM compatible with QFT, which is not easy 
to construct (for an attempt see e.g. \cite{nikQFTpilot,nikbook}).

{\em Bohmian interpretation 4}: 
This version of BM denies the existence of quark and electron trajectories
${\vec Q}_{\rm quarks\; \& \;electrons}(t)$, but accepts the
existence of as yet unknown {\em truly} fundamental particles with Bohmian trajectories 
${\vec Q}_{\rm truly \; fundamental}(t)$. 
The truly fundamental particles cannot be created and destroyed, 
so they are described by non-relativistic QM. 
This bypasses the hard problem of relativistic BM 
in the Bohmian interpretation 3. 
The Bohmian interpretation 4 is the version of BM that we actually propose in this paper.

\section{Summary}
\label{SEC6}

In this paper we introduced the notion of 
perceptibles, which are macroscopic entities that we observe directly. 
Since perceptibles are macroscopic, they are distinguished in the position space. 
The non-perceptibles such as wave function, atom, photon, etc. are
theoretical constructs that explain the perceptibles. 
Our main axiom for the perceptibles is that all perceptibles are beables. 
Essentially, it means that the Moon is there even when we do not observe it.  
However, there is no strict border between perceptibles and non-perceptibles, 
which suggests that microscopic positions are beables too. 
This is the motivation for introducing Bohmian mechanics (BM), 
according to which particles have trajectories. 

But {\em what} particles? BM only makes sense if the particle trajectories are 
trajectories of the fundamental particles. 
%
On the other hand, as discussed is Sec.~\ref{SEC5.1},
there are indications that the Standard Model particles might not be fundamental,
suggesting that BM should not be applied to them.
Moreover, the simplest version of BM makes a generic measurable prediction; it predicts
that the fundamental particles obey non-relativistic QM. 
Analogy with phonons indicates that fundamental non-relativistic QM 
may lead to non-fundamental relativistic QFT, which, in principle, 
bypasses the problem of relativistic BM without knowing the details of those 
hypothetic fundamental particles.

\section*{Acknowledgements}
The author is grateful to X. Oriols for discussions. 
This work was supported by 
the Ministry of Science of the Republic of Croatia,
by the European Union through the European Regional Development Fund 
- the Competitiveness and Cohesion Operational Programme (KK.01.1.1.06)
and by H2020 Twinning project No. 692194, ``RBI-T-WINNING''.

\end{document}